%% file: zhao2025nonreciprocal_rev2.tex
\documentclass[%
reprint,
 amsmath,amssymb,
 aps,
pre,
]{revtex4-2}

\usepackage{graphicx}
\usepackage{dcolumn}
\usepackage{bm}

\usepackage{ulem}
\usepackage{xcolor}

\newcommand{\Del}[1]{\textcolor{olive}{\sout{#1}}}
\renewcommand{\Del}[1]{}

\usepackage{soul}  

\begin{document}

\title{Nonreciprocal amplification toward chaos in a chain of Duffing oscillators}

\author{Luekai Zhao}
 \author{Bojun Li}%
\author{Nariya Uchida}%
 \email{nariya.uchida@tohoku.ac.jp}
\affiliation{%
 Department of Physics, Tohoku University, Sendai 980-8578, Japan
}%

\date{\today}

\begin{abstract}
{A chain of harmonic oscillators with nonreciprocal coupling exhibits
characteristic amplification behavior that serves as 
a classical analog of the non-Hermitian skin effect (NHSE).
We extend this concept of nonreciprocal amplification}  
to {nonlinear dynamics by employing} double-well Duffing oscillators arranged {in ring-structured units}. 
{The addition of units} induces bifurcations of attractors, 
{driving transitions} from limit cycles to tori, chaos, and hyper-chaos. 
{Unidirectional} couplings between units enable the decomposition of {attractors in phase space
into projected subspaces corresponding to each unit. }
In the chaotic regime, {amplitude saturation emerges,}
characterized by monotonically decreasing amplitudes within a unit
{-- in sharp contrast to} the increasing {profiles} seen in 
the linear NHSE. 
This work {uncovers} novel bifurcation {behavior resulting}
from the intricate interplay between {nonreciprocity} and nonlinearity.   
\end{abstract}

\maketitle


\section{Introduction} 
Physical systems with nonreciprocal couplings can {exhibit} 
exceptional phenomena 
such as nonreciprocal phase transitions~\cite{fruchart2021non} and odd elasticity~\cite{scheibner2020odd}.
{More broadly,} non-normality governs the stability of ecosystems through transient amplification~\cite{neubert1997alternatives,asllani2018topological,muolo2019patterns}.
{The non-Hermitian skin effect (NHSE) has recently emerged as a 
striking {manifestation}
 of nonreciprocal interactions in physical systems. 
It refers to a phenomenon {in which} small disturbances at the boundary cause 
the system’s wave-like modes (or eigenstates) to become strongly concentrated at one edge. 
Instead of being uniformly distributed, the amplitudes {accumulate} 
near a boundary due to asymmetries in how signals or energy propagate through the system~\cite{zhang2022review}.} 
The proliferation of experimental 
{results}~\cite{zhang2024construction,xiong2024tracking,zhang2021observation,helbig2020generalized,ghatak2020observation,liang2022dynamic,brandenbourger2019non} {across various} platforms further {highlights}
the significance of NHSE. 
{Since} its origin {lies} in linear system{s}, 
{a} natural {next step is to
explore its generalization to nonlinear systems}. 
A typical system used to study nonlinear extension{s} of NHSE is the discrete Schr{\"o}dinger equation with Kerr nonlinearity~\cite{yuce2021nonlinear,ezawa2022dynamical,jiang2023nonlinear}. 
{Under certain conditions, l}ocalized stationary solutions 
{analogous} to
eigenstates in linear systems are analytically derived~\cite{yuce2021nonlinear}.
{In addition, numerical studies of the} 
quench dynamics {have identified multiple} skin states~\cite{ezawa2022dynamical}, {and 
a}nalytical solutions {known as} skin discrete breathers {have been shown to exhibit double-exponential decay}~\cite{jiang2023nonlinear}.
Furthermore, {nonlinear} Su-Schrieffer-Heeger-type {models} 
{have been developed to study unique properties of localized solutions~\cite{lang2021non,manda2024insensitive}}.

Although most of the nonlinear extensions of NHSE have {focused on} quantum systems, {the effect is fundamentally rooted in dynamical systems.
Localized solutions to the dynamical equations exhibit nonreciprocal amplification -- namely a spatial growth in 
oscillation amplitude along the direction of nonreciprocity~\cite{wen2023acoustic,galiffi2019broadband,lin2019nonreciprocal}. 
While nonreciprocal amplification is a hallmark of linear NHSE, it represents a more general mechanism that can also arise in nonlinear systems with open boundaries.}

As a paradigmatic prototype of the nonlinear dynamic{al} system, 
Duffing oscillators {have been widely studied, and are applicable to}
many physical {contexts}~\cite{kovacic2011duffing, virgin2000introduction, jones2001duffing, salas2021duffing}. 
{A} single Duffing oscillator with time-dependent driving 
{can generate} chaotic attractors~\cite{ravichandran2007homoclinic},
{and
coupled Duffing oscillators can give rise to even richer dynamics.}
{In} a dissipative{,} nonautonomous system, 
two Duffing oscillators driven by {a periodic} external force {exhibit} complex {bifurcation routes} 
to chaos~\cite{kenfack2003bifurcation}. 
In autonomous systems, {however,} dissipation {makes}
bifurcations to chaos {more
difficult}~\cite{sabarathinam2013transient,sabarathinam2015transient}, and requir{ing careful} structural design.
A ring of {three or more Duffing oscillators 
coupled unidirectionally overcomes this}
challenge~\cite{perlikowski2010routes}, {exhibiting}
hyperchaos via various {bifurcation types}.
{In such systems, chaotic rotating waves can also arise}
from spatiotemporal symmetry, {forming a basis}
for {studies on}
coexisting rotating waves in larger rings~\cite{barba2023dynamics}.
{Beyond} linearly coupled {systems},
{nonlinearly coupled} Duffing oscillators
{led to} 
multi-spiral chaos~\cite{balaraman2023coexisting}, 
{exact solutions~\cite{lenci2022exact},
and dependence of bifurcation routes on the number of degrees of 
freedom}~\cite{musielak2005chaos}. 

{In this work, we explore bifurcation routes of attractors with
localized dynamical patterns 
in a chain of linearly coupled autonomous Duffing oscillators.
Nonlinearity and dissipation give rise
to localized patterns that differ fundamentally  from
those found in linear NHSE.
Due to dissipation,
trajectories starting from different initial conditions within 
the same basin of attraction converge to a common asymptotic state,
allowing us to characterize global dynamics without exhaustive sampling.
The chain is constructed by connecting 
ring-structured units via unidirectional couplings.
Long-range backward couplings within each unit compensate for energy loss 
and support nontrivial attractors and their bifurcations.
The unidirectional inter-unit couplings
permit the decomposition of attractors in 
high-dimensional} phase space into independent subspace manifolds.
{As the number of units increases, an unconventional bifurcation route 
emerges -- one governed more by system size than coupling strength.
In the absence of damping, nonlinear cubic terms dominate over 
the nonreciprocal couplings and destroy the localized patterns.
Dissipation suppresses amplitude growth and restores
nonreciprocal amplification in the velocity profiles.
Consequently,  a saturated amplitude profile appears beyond a certain unit,
where dissipation and backward coupling jointly suppress further growth. 
This results in a decline in velocity amplitudes along the chain,
contrasing with the monotonic growth seen in linear NHSE.
}
 
\section{Model}
\begin{figure}[htbp]
    \includegraphics[width=\linewidth]{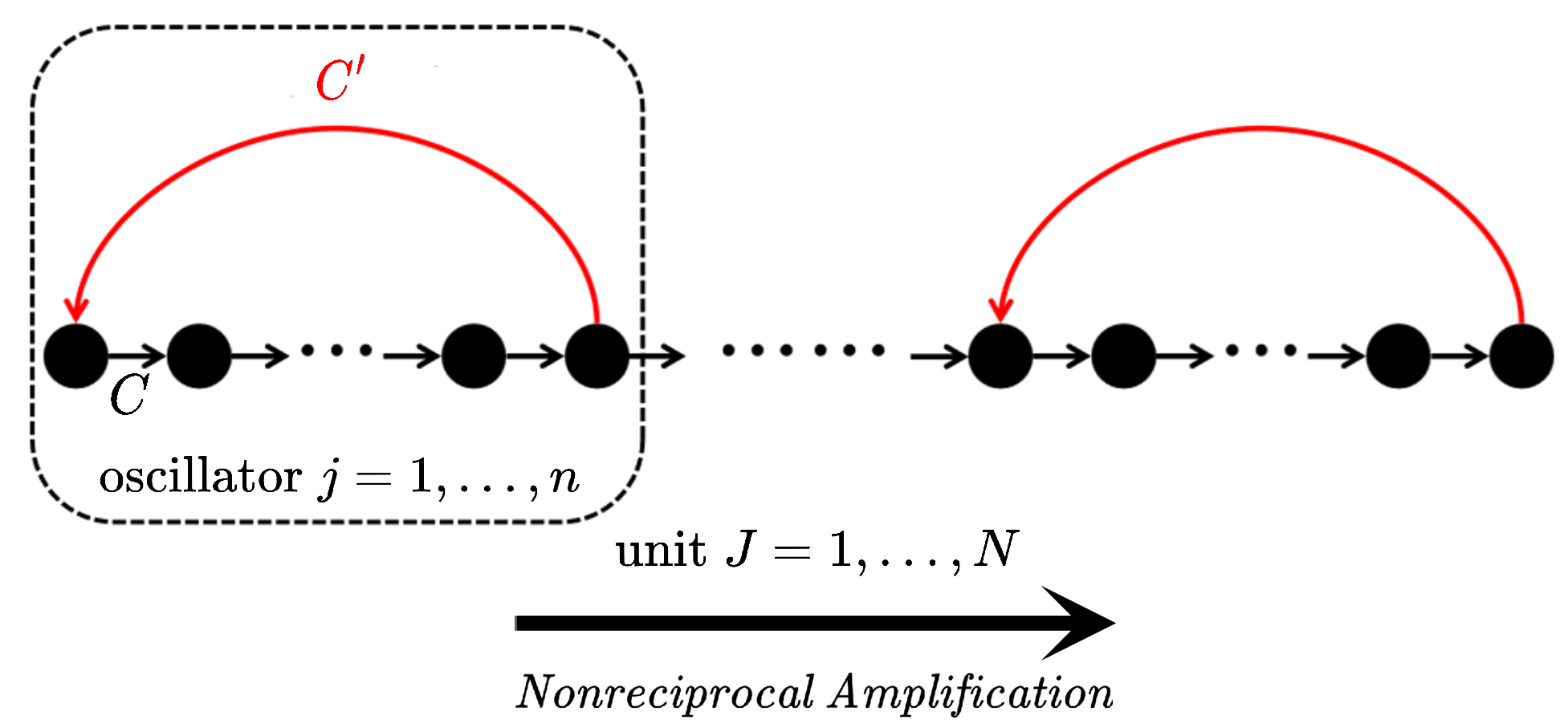}
    \caption{{Schematic of a one-dimensional chain of 
non-reciprocally coupled} Duffing oscillators. 
{The system consists of $N$ units, each 
containing $n$ oscillators.
The first unit is enclosed by the dashed black rectangle.
Solid black circles represent dissipative double-well Duffing oscillators, 
black arrows indicate nearest-neighbor couplings $C$,
and red arrows denote long-range compensation couplings $C^\prime$.
} 
}    
\label{fig:1}
\end{figure}

The dynamics {of} {a} one-dimensional chain of linearly nonreciprocally coupled double-well Duffing oscillators is {govened by} the {following} second{-}order ordinary differential equations (also see \figurename~\ref{fig:1}): 
\begin{widetext}
\begin{equation}
    \dot{x}_{J,j}=v_{J,j}, \quad \forall J,j
\end{equation}
\begin{equation}
    \dot{v}_{J,j}=\left\{\begin{matrix}\begin{array}{l l l}
 - k x_{J,j}^3 + \kappa x_{J,j} + C (x_{J,j-1} - x_{J,j}) 
    - \gamma v_{J,j}, & j \neq 1\\
  - k x_{J,j}^3 + \kappa x_{J,j} + C (x_{J-1,n} - x_{J,j}) 
    + C^\prime (x_{J,n} - x_{J,j}) - \gamma v_{J,j}, & j = 1 \text{ and } J \neq 1 \\ 
  - k x_{J,j}^3 + \kappa x_{J,j} + C^\prime (x_{J,n} - x_{J,j})
    - \gamma v_{J,j} ,& j = 1 \text{ and } J = 1
\end{array}
\end{matrix}\right.
\end{equation}
\end{widetext}
where {$J = 1, 2, \ldots, N$ is the index of the units, 
and $j = 1, 2, \ldots, n$ is the index of oscillators within each unit.}
{$x_{J,j}$ and $v_{J,j}$ represent the} displacement and velocity of each oscillator{, respectively. The parameters} $k$, $C$, $C^\prime$, {and} $\gamma$ denote the coefficients of nonlinearity,
nearest-neighbor couplings (black arrows), long-ranged compensation couplings (red arrows) and damping.
$\kappa$ is {essential for} creating {the} double-well potential{,} 
and the resulting three {prominent} fixed points are located at (\(x_{J,j}=0,v_{J,j}=0\)) and (\(x_{J,j}= \pm \sqrt{\kappa / k},v_{J,j}=0\)) in the phase space for \(\forall J,j \). 
The {forward coupling} $C$ 
between every pair of nearest-neighbor oscillators {induces 
nonreciprocal amplification, which is defined as an increasing tendency of 
the velocity amplitudes of the oscillators from left to right.}
{On the other hand,} the {backward coupling} $C^\prime$ 
compensate for the energy loss of the leftmost oscillator in each unit. 

As the system {is dissipative and lacks} external driving, 
$C^\prime$ is {essential for driving the first oscillator of the chain
and} the existence of attractors other than fixed points. 
{In the nonreciprocal limit  $C^\prime=0$,
the leftmost oscillator ($J=j=1$) 
is decoupled from the system and falls exponentially fast to a fixed point. 
The resulting loss of energy transfer through the unidirectional coupling ($C$) 
causes 
the other oscillators to reach the fixed points one by one, from left to right.}
On the other hand, when $C^\prime$ is {increased,}
the {nonreciprocal amplification} resulting from $C$ 
 is weakened and distorted by $C^\prime$ in the opposite direction.
As {we will show} later,
different attractors {emerge along a classical bifurcation} route described in~\cite{perlikowski2010routes}
{as the ratio $C^\prime / C$ is varied.}
Throughout this {study}, we fix $\kappa/k=1/2$ and 
$\gamma=0.4$, {while varying $k$, $C$, and $C'$ 
such that the ratio $C^\prime/C$ lies between 0 and 1.
Parameter values are provided in the corresponding 
figures and captions.
For numerical integration of Eqs.(1), (2), we used the fourth-order 
Runge-Kutta method with the time increment $\Delta t=0.01$.
}

\section{Results}
{We begin by discussing} the dynamics of our model, 
first {at the level of a single} unit and then {for} 
the one-dimensional chain {composed of multiple} units.
We {employ} numerical {methods to obtain}
phase portraits, the Lyapunov spectrum~\cite{benettin1980lyapunov,sandri1996numerical} and  
{spatial profiles of the velocity amplitudes}. 
\begin{figure*}[htbp]
    \centering
    \includegraphics[width=0.65\textwidth]{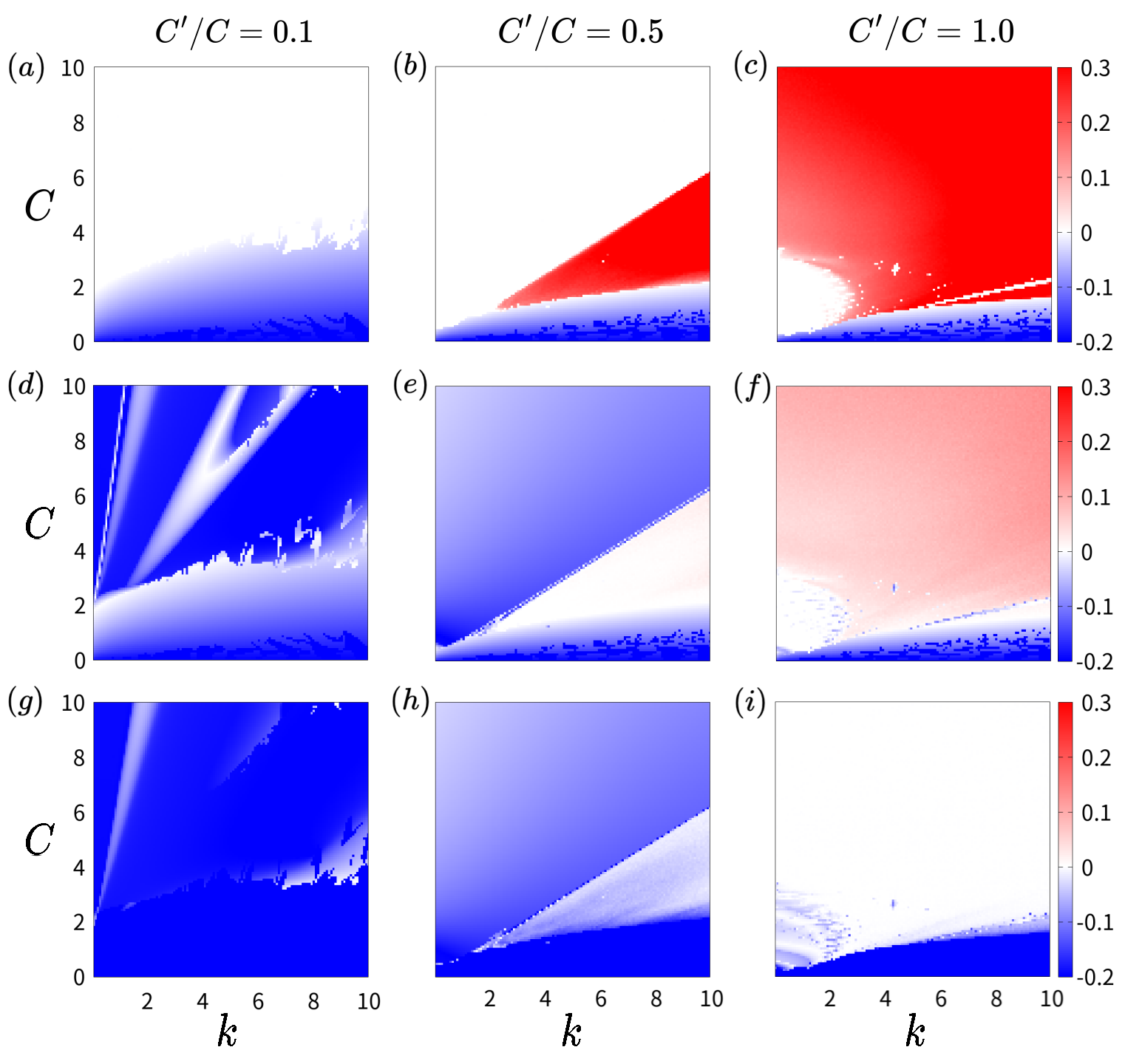}
    \caption{The first three {largest Lyapunov exponents (LLEs)} in 
{the} $k$-$C$ parameter space. 
{Panels} (a)-(c) {show} the first LLE, (d)-(f) the second LLE, (g)-(i) the third LLE,
for $C^\prime/C=0.1, 0.5, 1.0$ from left to right.
{The sign of the first LLE distinguishes chaotic (positive, red),
periodic and quasi-periodic  (zero, white), 
and fixed points (negative, blue) attractors.
Hyperchaos is defined by a positive second LLE.
}
The Lyapunov {spectra} are {computed} 
over {the} time interval from {$20,000$  
to $50,000$ time units.}
}
    \label{fig:2}
\end{figure*}
\subsection{Dynamics of {a single} unit}
As a starting point, we delve into the mechanism by which a single unit 
{exhibits} 
bifurcations and {investigate} the routes to chaos under the influence of the compensation ratio $C^\prime / C$.

{Although the unit with the minimal size ($n=2$)} has been 
{studied} in the previous {work}~\cite{perlikowski2010routes}, 
here we explicitly demonstrate that there are only fixed points for two oscillators {using an} analytical approach {based on the} 
Lyapunov function.
The Lyapunov function, which also {serves as} the Hamiltonian 
{in the} conservative {case}~\cite{zhou2021most}, 
for a single unit consisting of two oscillators is {given by}
\begin{gather}
 L(\bm{x},\bm{v}) = \frac{1}{2} \bm{v}^2 + \frac{k}{4}(x_1^4 + x_2^4 ) - \frac{\kappa}{2} \bm{x}^2 + \frac{C}{2} (x_1-x_2)^2 .
\end{gather}
{Here, \(\bm{x}=(x_1,x_2)\), \(\bm{v}=(v_1,v_2)\), 
and we set $C^\prime / C$} to 1. 
In general, there are two global minima { \((\bm{x,\bm{v}})_{min}=(\pm \sqrt{\kappa / k},\pm \sqrt{\kappa / k},0,0)\)} 
and one local maximum{ \((\bm{x,\bm{v}})_{max}=(0,0,0,0)\); see Appendix A for details.} 
{Thus, fixed points are}  
the only possible attractors {for} $C^\prime=C$, 
{which gives maximal compensation.
Therefore, we also obtain only fixed points for $C^\prime < C$.}
{This result indicates} that the {backward coupling} $C^\prime$ between nearest-neighbor oscillators fails to compensate for the energy loss. 
{This is because 
the forward coupling $C$ from the left oscillator to the right one
lacks an amplification mechanism along the chain,  
and as a result, the right one does not gain sufficient energy to feed back 
to the left.}
\begin{figure*}[htbp]
    \centering
    \includegraphics[width=1.0\textwidth]{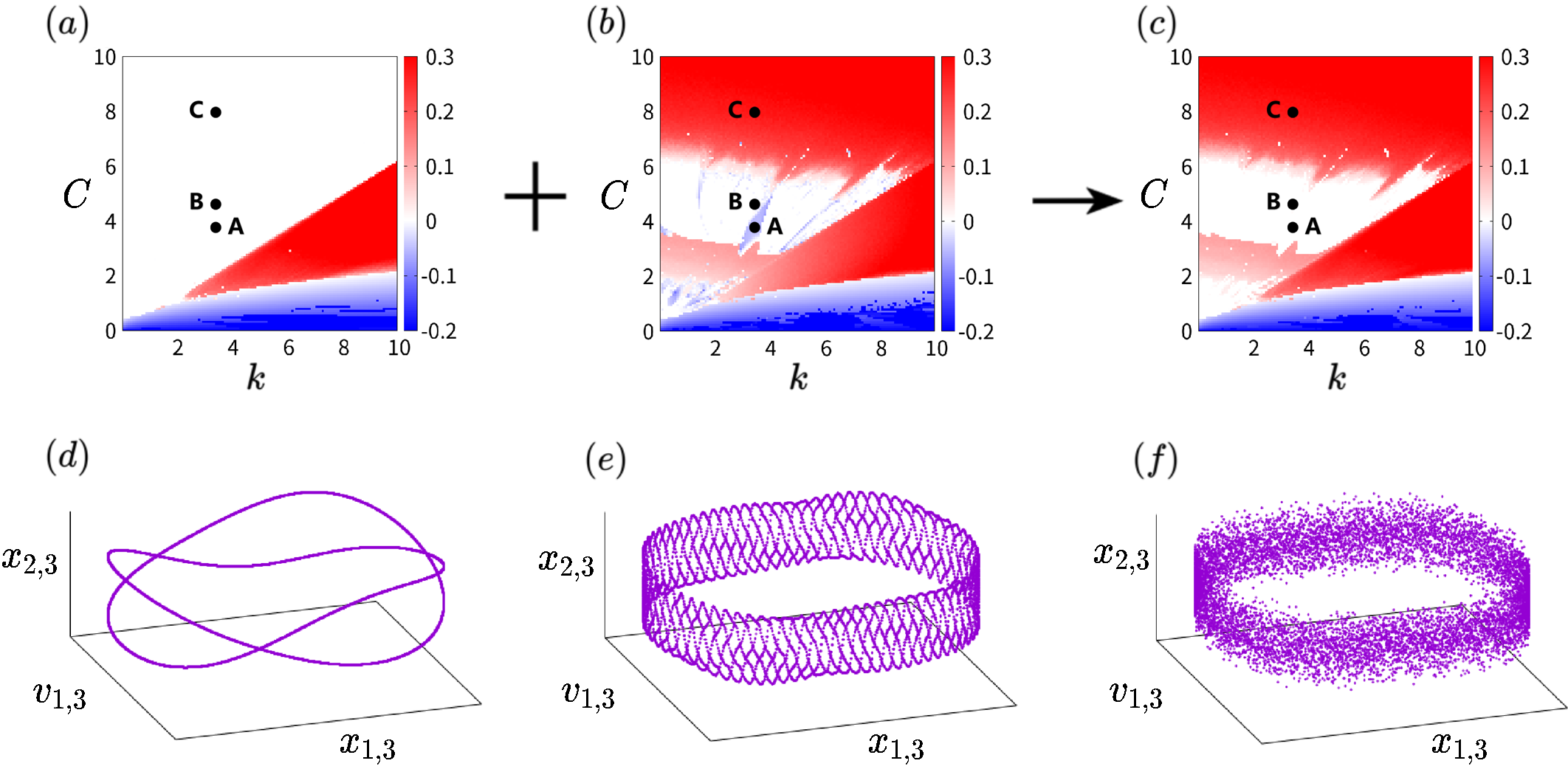}
    \caption{Illustration for the synthesis of attractors in 
{high-dimensional phase space}. 
    {(a) First LLE for a single unit with $(N,n)=(1,3)$. 
    (b) First LLE for the second unit ($J=2$) in a two-unit system with $(N,n)=(2,3)$. 
    (c) First LLE for the two-unit system with $(N,n)=(2,3)$.}
{ (d-f) Phase portraits for three representative points:} 
(d) {Point} \textbf{A} $(k=3.4, C=3.7)${;}  
(e) {Point} \textbf{B} ($k$=3.4, $C$=4.7){;}
(f)  {Point} \textbf{C} $(k=3.4, C=8)${; all for $(N,n)=(2,3)$. 
   The} projection plane is spanned by $x_{1,3}$ and $v_{1,3}$. 
    Phase portraits are {computed over the time interval from} 45,000 to 50,000 time units. {In all panels, $C^\prime / C =0.5$.}
}
    \label{fig:3}
\end{figure*}
This {motivates our subsequent study,} where we {extend}
the unit to {increase} the strength of the rightmost oscillator 
{through} nonreciprocal amplification. 
{The resultant strong compensation for the leftmost oscillator
allows the entire unit to evolve into
periodic, quasi-periodic, and chaotic dynamical patterns.} 

{The behavior of the unit with $n=3$ has been studied only for uniform couplings ($C^\prime=C$)~\cite{perlikowski2010routes, barba2021dynamics}, where 
Hopf bifurcations are obtained by varying the coupling strength.}
{However, our focus here} 
is not only on bifurcations associated with {various} attractors, 
but also on the {nonreciprocal amplification within} the unit, 
defined as the increasing tendency of {velocity} amplitude
from left to right.
{Note that, for $C^\prime=C$, the unit is regarded as a uniform ring
or a linear chain with the periodic boundary conditions,
for which we cannot expect nonreciprocal amplification.}
Therefore, we {mainly examine the case $C^\prime < C$, 
where}
{nonreciprocal amplification is observed within the unit. }
\figurename~\ref{fig:2} {shows 
the three largest Lyapunov exponents (LLEs)
for various values of $C^\prime/C$.}
An effective way {to identify} different {types of} attractors 
{is to consider the signs of the triplet}
(first LLE, second LLE, third LLE),  shown by colors in Fig.~2;
 $\text{LLE} > 0$ (red); $\text{LLE} = 0$ (white); $\text{LLE} < 0$ (blue).
{The} regions in the $k$-$C$ space are classified as {follows:}
fixed points (blue, blue, blue), limit cycle (white, blue, blue), 
2D torus (white, white, blue), 3D torus (white, white, white), 
chaos (red, white, blue), hyperchaos (red, red, white).
{(Note that hyperchaos is defined by a positive second LLE.)}
Compared to \figurename~\ref{fig:2}(c), the {corresponding} regions 
in \figurename~\ref{fig:2}(a) and \figurename~\ref{fig:2}(b) 
{do not exhibit} 
chaotic behavior {due to insufficient} compensation 
{by} $C^\prime$.
Next, for each {fixed} value of $k$, we {analyze how the
system transitions to chaos as the coupling strength $C$ increases.}
In \figurename~\ref{fig:2}(a)-(c), {a} Hopf bifurcation {occurs}
for all $k$ as $C$ {crosses the boundary} of the blue region.
For {larger} nonlinear stiffness $k$, the Hopf bifurcation {requires} stronger coupling {strengths} $C$, and {chaotic dynamics emerges more readily.}

It is noteworthy that the bifurcations {are driven by} the {nonreciprocal amplification within}  the unit.
The rightmost oscillator{, having the highest amplitude
due to nonreciprocal amplification}, 
is coupled to the leftmost oscillator{, which has} the lowest amplitude. 
The amplitude gap between the {right}most and {left}most oscillators is {large} enough to {prevent} 
the leftmost oscillator {from decaying by dissipation, 
thereby enabling}  bifurcations {of} the attractors.
A contrasting model with the same ring structure but reciprocal nearest-neighbor couplings {is presented in Appendix B}, where 
the {system does not exhibit} bifurcations.

Moreover, it can be inferred that {a} unit with more oscillators 
{exhibits} stronger {nonreciprocal amplification} and requires less compensation coupling {to follow} similar routes to chaos.
In other words, due to {nonreciprocal amplification}, a bifurcation toward chaotic attractors depends critically on the unit size $n$.
The resultant attractors reside in the phase space of all oscillators in a unit.

\subsection{Attractor synthesis and nonreciprocal amplification in a chain}

 {Next, we connect} several units via unidirectional couplings 
{to} obtain a one-dimensional chain. 
{We start from a chain} of $(N, n)=(2,3)$, 
which is the minimal system {that exhibits} bifurcations
{as previously demonstrated~\cite{perlikowski2010routes}.} \figurename~\ref{fig:3}(c) shows the first LLE of this {two-unit} system. 
{
The difference from the single-unit case (FIG. 3(a)) 
originates solely from the second unit ($J=2$).
FIG. 3(b) shows the first LLE in the subspace spanned by 
the positions and velocites of the newly added unit.
We can reproduce FIG.~3(c) by 
taking the pointwise maximum of the LLEs shown in FIG. 3(a) and FIG. 3(b).
}
{Note that such decomposition is generally 
impossible in a chain of oscillators that are mutually coupled.
It arises solely from the unidirectional coupling, by which the newly added unit is driven by the old unit ($J$=1) without affecting it 
in return.
Thus, the attractors in the phase space of the two-unit system can be regarded as a synthesis of the attractors from the original and the newly added units.
When projected onto the subspace corresponding 
to the old unit, the attractors of the two-unit system 
remain identical to those of a single unit.
In contrast, the projection onto the subspace 
of the newly added unit reveals
additional dynamical features.}
{We demonstrate this using
three representative examples of attractor synthesis obtained at 
the parameter points \textbf{A},\textbf{B},\textbf{C} 
in \figurename~\ref{fig:3}.} 
From the phase portraits for $(N,n)=(2,3)$, 
we {find that all of them exhibit} limit cycles 
in the projection plane{, identical to the attractors of the original unit shown} in \figurename~\ref{fig:2}(b){; see also} \figurename~\ref{fig:2}(e). 
The newly added unit {exhibits distinct attractor trajectories 
outside} the projection plane, {consistent} with the first LLE{s indicated by the three points}  in \figurename~\ref{fig:3}(b). 
Altogether, the {full system exhibits synthesized attractors 
in higher-dimensional phase space{,
meaning that the global dynamics can be viewed as a combination of the attractors from individual units, each driven unidirectionally by its upstream neighbor.}

\begin{figure}[htbp]
\includegraphics[width=1.0\linewidth]{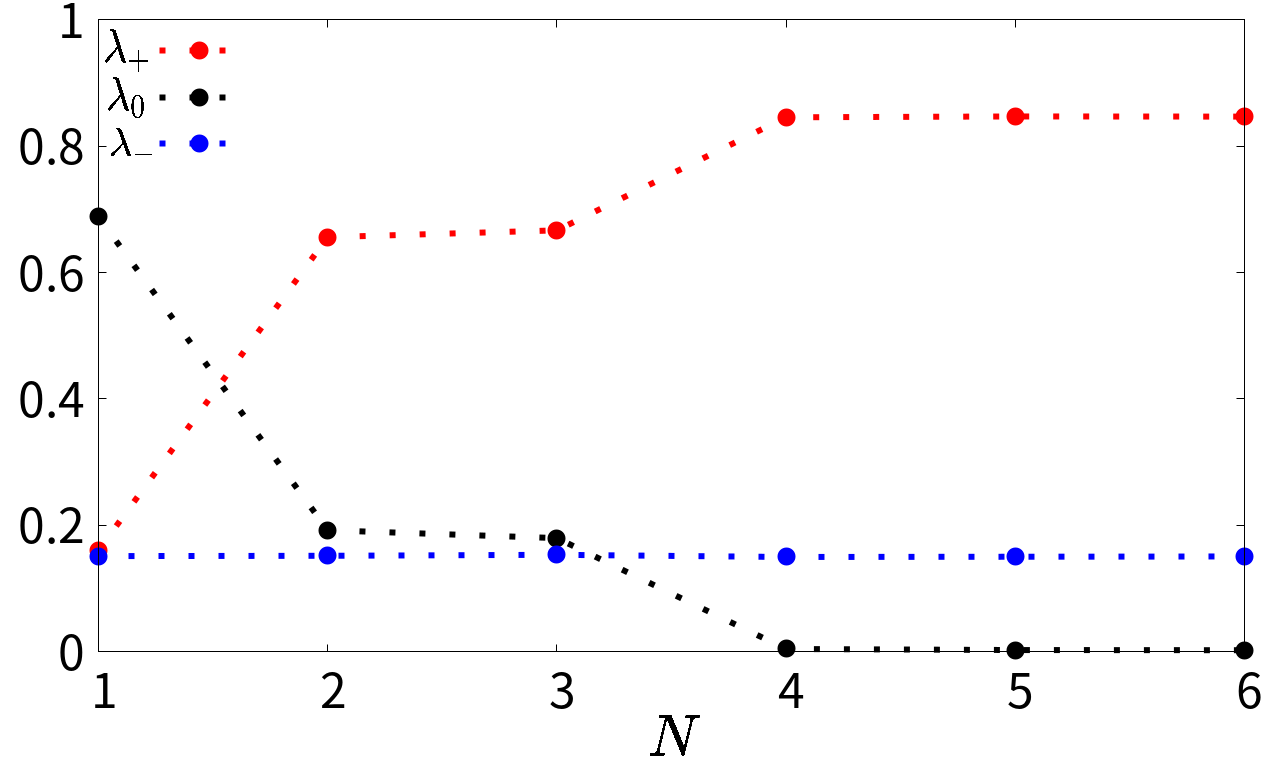}
\caption{{Area fractions of  the chaotic (red), 
periodic and quasi-periodic (black), and fixed-point (blue)
regions in
the $k$-$C$ space [as in \figurename~\ref{fig:2}(b)],
plotted as a function of the number of units $N$. }
The three regions are {classified according to the value
of the first LLE:
$\lambda _+$ (positive), $\lambda _0$ (zero), 
and $\lambda _-$ (negative),
respectively. 
See Appendix C for the precise definition of the
intervals.
The total area of the $k$-$C$ space considered 
is same as in \figurename~\ref{fig:2}(b), with
$0 < k \le 10$ and $0\le C \le 10$. Here, we fix $C^\prime / C$=0.5.}
}
\label{fig:4}
\end{figure}

\figurename~\ref{fig:4} {illustrates
the area fractions of the chaotic,  periodic and quasi-periodic, 
and fixed-points regions in the $k$-$C$ space 
as a function of the number of units $N$. 
The dominance of fixed points 
-- characterized by negative first LLE{s} --
remains nearly unchanged as the system length increases.
This behavior can be interpreted as a consequence 
of unidirectional coupling:
if the dynamics of one unit converges to a fixed point, 
(i.e.,  a stationary state), it cannot transfer energy to the 
subsequent unit. As a result, the dynamics of the next unit
also converges to a fixed point. }
{Periodic} and quasi-{periodic} attractors 
{-- characterized by vanishing first LLEs --} 
are gradually transformed into chaotic attractors
{with positive first LLEs} 
as more units are added.
{Chaotic attractors almost completely dominate when
the system reaches $N=4$.
This reveals a new and simple route to chaos:
increasing the number of units in the system
without altering the parameters in the dynamic equations.
}

For the longer chain system ($N=20$), we examine 
{nonreciprocal amplification,} 
defined {as an} 
increasing tendency of the time-averaged velocity amplitude 
from the first unit {($J=1$) to the rightmost unit ($J=N$).}
{The} time-averaged velocity amplitude for each unit $\langle \overline{|v|}_{J} \rangle_t$ and each oscillator $\langle |v_{J,j}| \rangle_t$ are defined by
\begin{equation}
    \langle \overline{|v|}_{J} \rangle_t = \frac{1}{\Delta t \cdot n} \sum_{t}\sum_{j=1}^{n}|v_{J,j}(t)|,
\end{equation}
\begin{equation}
    \langle |v_{J,j}| \rangle_t = \frac{1}{\Delta t} \sum_{t}|v_{J,j}(t)|.
    \label{eq:5}
\end{equation}
{The amplitude} $\langle \overline{|v|}_{J} \rangle_t$ 
{characterizes nonreciprocal amplification
at the level of the entire chain, whereas} $\langle |v_{J,j}| \rangle_t$ {captures nonreciprocal amplification within individual units.
Since fixed points cannot exhibit nonreciprocal amplification,
we focus on parameter values within} 
the white region of \figurename~\ref{fig:2}(b). 

\begin{figure}[htbp]
    \includegraphics[width=1.0\linewidth]{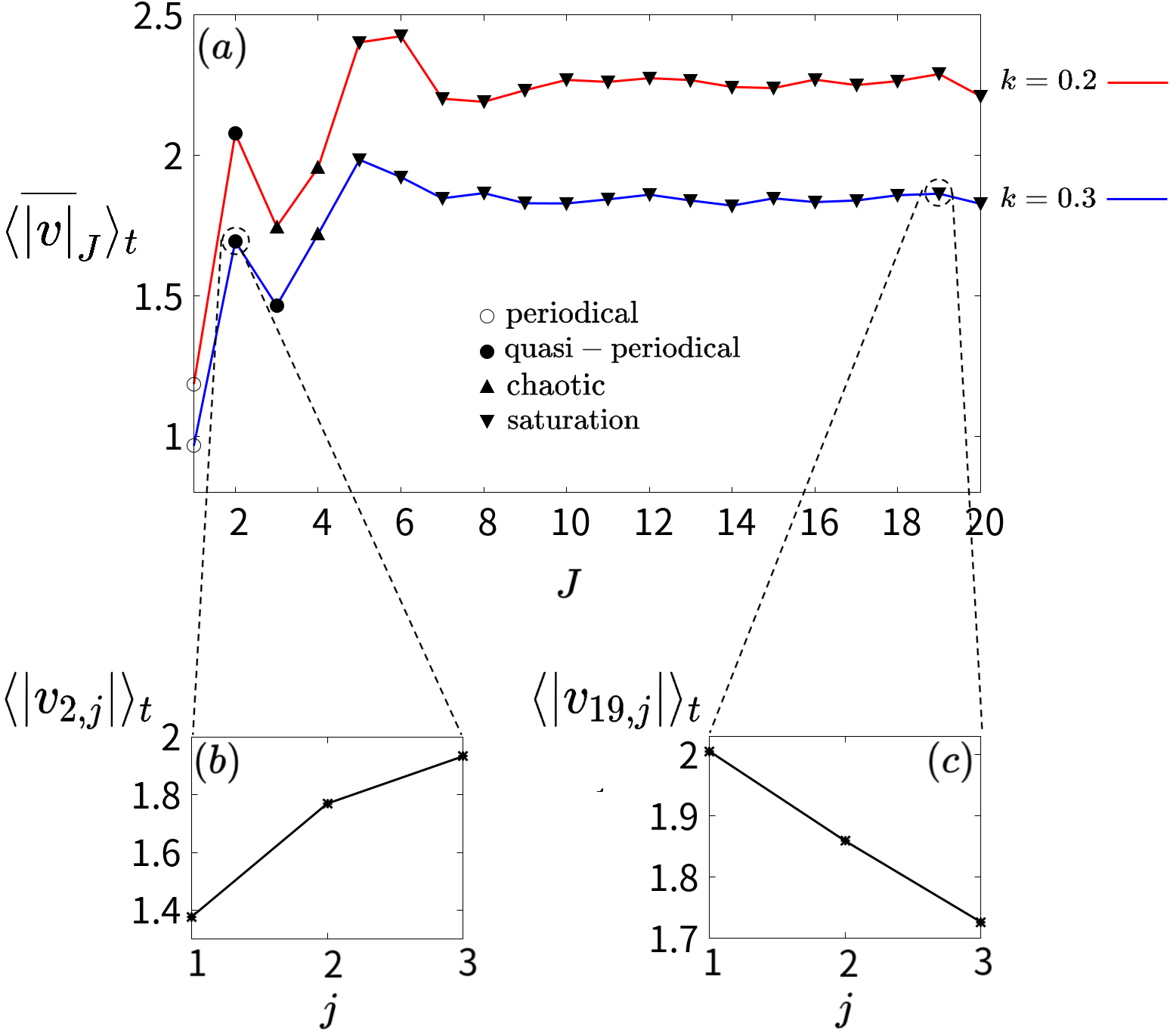}
    \caption{{Spatial profiles of the time-averaged velocity amplitides.} 
(a) {Time-averaged total} velocity amplitude 
{for} each unit{,} 
$\langle \overline{|v|}_{J} \rangle_t${, plotted against the} unit index $J$. 
{Hollow circles, solid circles, solid triangles, and inverted solid triangles indicate the types of attractor in each unit. 
}
(b) {Time-averaged} velocity amplitude{s} of oscillators 
in the second unit{,} 
$\langle |v_{2,j}| \rangle_t${, plotted against the} oscillator index $j$. 
(c) {Same as (b), but for} the nineteenth unit{,} 
$\langle |v_{19,j}| \rangle_t$. 
{All time averages are computed over the interval} 
from 30,000 to 50,000 time units. {We set $C=0.9$ and $C^\prime / C= 0.5$}.}
    \label{fig:5}
\end{figure}
\figurename~\ref{fig:5}(a) {compares the velocity amplitude profiles
for different values of nonlinear stiffness} $k$. 
Weaker stiffness $k$ {leads to}  higher velocity amplitudes.
{Moreover,} since {the attractor in unit $J$ induces 
the dynamics in $(J+1)$} via unidirectional {coupling}, 
bifurcations to chaos {can be visualized} by examining the attractors in each unit from left to right in \figurename~\ref{fig:5}(a).
{A characteristic feature of the amplitude profile} 
is that {it saturates beyond} a certain unit {along} 
the one-dimensional chain. 
{In contrast, the amplification continues indefinitely with increasing system size in linear systems}~\cite{okuma2020topological}.
{In our case, the} saturation arises from the nonlinear and dissipative nature of the dynamics.
{In this context, saturation refers to a sustained chaotic pattern that begins at a specific unit and persists throughout the remaining chain.}
{
This pattern is characterized by a monotonically decreasing
velocity {amplitude within each} unit, 
{as shown in} \figurename~\ref{fig:5}(c).
{This descending pattern, shown} in \figurename~\ref{fig:5}(c), 
is found to be robust {across} a wide range of parameter values. 
Moreover, the ratio $C^\prime / C$ affects the rate of decay, 
as discussed in {Appendix D}.}
The {observed} saturation {primarily arises from} 
the interplay {between} Duffing nonlinearity and damping,
{which together impose} an upper bound on the growth of attractor size.
To {clarify} the effect of saturation, we {examine} the time-averaged velocity amplitude{s}  in two {adjacent} units. 
In \figurename~\ref{fig:5}(b), {nonreciprocal amplification} is still observed in the second unit, {whereas in} \figurename~\ref{fig:5}(c), the {descending pattern} dominates in the nineteenth unit.
In the second unit, the oscillators have not {yet reached} saturation, allowing the attractor size {to grow}. {In contrast, in}
the nineteenth unit, saturation suppresses {attractor} growth, 
while the compensation coupling {enhances} the amplitude of 
the first oscillator, resulting in the formation of the {descending pattern}.
In {summary}, as the site index $J$ {increases, each} 
unit approaches saturation and tends to 
lose {nonreciprocal amplification within the unit}.
{As the system size increases, chaotic dynamics become dominant, 
as shown in \figurename~\ref{fig:4}, 
while saturation is observed in  \figurename~\ref{fig:5}(a). 
These observations together confirm that saturation occurs within 
the chaotic regime.}
For instance, {parameter values of $k$ and $C$ within} 
the red region of \figurename~\ref{fig:2}(b) 
{bring} the single unit close to saturation,
{thereby leading to the emergence of} chaos. 
 
\section{Discussions}
{In this work, we have investigated the rich bifurcation behavior of attractors induced by nonreciprocal coupling.}
{First,} we analytically 
{demonstrated the absence of bifurcations from fixed points 
in the case of a unit with $n=2$.}
Then, by {tuning} the compensation ratio, we 
{found} {nonreciprocal amplification in} 
a single unit and {obtained distinct} bifurcation diagrams. 
{Importantly, {nonreciprocal amplification} is {essential for inducing} bifurcations in the presence of compensational coupling}.
Next, we {turned} to the one-dimensional chain system {composed} of {multiple} units.
{Due} to the unidirectional couplings, the {phase spaces of
upstream} units are {effectively decoupled,} 
{enabling} the synthesis of {high-dimensional} attractors.
{Finally, nonreciprocal amplification was demonstrated
in conjunction} with bifurcations to chaos {as the number of units is increased.}
{In the chain system, amplification is sustained up to the point of saturation, beyond which the system} 
transitions into a chaotic descending pattern.
{The dissipative and nonlinear nature of our model 
{may help bridge theoretical studies on}
nonreciprocal interactions {with} real-world problems{~\cite{wen2023acoustic,galiffi2019broadband,lin2019nonreciprocal,tripathi2024nanoscale,wu2022interplay}}.}
{For instance, our model aligns naturally with non-normal network architectures, which are known to capture the structure and dynamics of various real-life networks~\cite{asllani2018structure}. 
{Quantitative measures of non-normality
have been developed and used to assess the system's stability~\cite{trefethen2020spectra}.
To check their validity in our model, we have computed 
the numerical and spectral abscissas of the Jacobian matrices 
at the relevant fixed points. The numerical abscissa remains positive, indicating transient growth
due to non-normality, while a sign change of the spectral abscissa 
predicts the onset of Hopf bifurcation. 
These measures thus capture essential aspects of local linear dynamics, although a full understanding of global bifurcation behavior—particularly the transition to chaos—will require more comprehensive analytical tools.
}
Furthermore,  transitions between equilibria—induced by transient amplification of disturbances—underscore the significance of nonreciprocal amplification, which serves as the driving mechanism for bifurcations in our system.} 

{Many questions remain to be addressed 
in future work. 
First,
the coexistence of attractors and the structure of 
their basins of attraction} basins are not yet 
{fully understood (see Appendix E). 
For instance, according to~\cite{barba2023dynamics}, additional fixed points exist beyond the three prominent ones when the coupling strength is relatively small.
Second,} from \figurename~\ref{fig:2}(b), the compensation from $C^\prime$ is {insufficient, leading to
re-entrant transitions} from chaos to {periodic or quasi-periodic} attractors. 
This {points} to an underlying phenomenon
{known as} transient chaos~\cite{lai2011transient}. 
{Third, the} rotating wave dynamics observed 
{in single-unit systems~\cite{perlikowski2010routes,borkowski2015experimental} could be very different
may differ significantly in the context of
one-dimensional chains.} 
Furthermore, as observed in \figurename~\ref{fig:5}(a), the increasing tendency {in} the first {few} units is not monotonic, which is associated with bifurcations leading to chaos and warrants further investigation.
\Del{As an analytical tool, measures of non-normality~\cite{trefethen2020spectra} can be applied to our system, providing a promising avenue for uncovering the deeper influence of non-normality on bifurcations leading to chaos.}
{As an extention of the present study, 
various models of coupled oscillators capable of 
generating spatiotemporal chaos --
such as the Brusselator model~\cite{castelino2020spatiotemporal} -- 
can be considered.
The synchronization dynamics of Brusselators 
with  non-normal coupling 
has been addressed in Ref.~\cite{muolo2020synchronization},
and its extension to the chaotic regime presents
an interesting direction for future research.
The linear coupling term can be also modified to 
a nonlinear one, as considered in Ref.~\cite{balaraman2023coexisting}.
Such a modification aims} to overcome the 
{limitations} of nonreciprocal amplification 
{arising from} linear couplings 
and to enable bifurcations in a unit {consisting} 
of two oscillators.    

{In summary, this work extends the concept of 
nonreciprocal amplification to 
nonlinear dynamical systems, providing new insights into 
the interplay between nonreciprocity in 
one-dimensional systems
and bifurcations toward chaos in high-dimensional phase space.
}

\begin{acknowledgments}

We thank Jiang Hui for useful discussions.

\end{acknowledgments}

\appendix

\renewcommand{\theequation}{A\arabic{equation}}
\setcounter{equation}{0}
\renewcommand{\thefigure}{A\arabic{figure}}
\setcounter{figure}{0}

\section{Lyapunov function}

Here, we analytically study the attractors for $(N,n)=(1,2)$, {with} $C^\prime = C$ using {the} Lyapunov function{:}
\begin{equation}
\begin{split}
 L(x_1,v_1,x_2,v_2) = \frac{1}{2} (v_1^2+v_2^2) + \frac{k}{4}(x_1^4 + x_2^4 ) \\- \frac{\kappa}{2} (x_1^2+x_2^2) + \frac{C}{2} (x_1-x_2)^2 > - \infty.
\end{split}
\end{equation}
{The time derivative} of $L(x_1,v_1,x_2,v_2)$ {is} 
\begin{eqnarray}
\frac{\mathrm{d} L}{\mathrm{d} t} & = & -\gamma (v_1^2+v_2^2) \le 0.
\end{eqnarray}
{This derivative} is zero {only on the two-dimensional}  subspace $(x_1,0,x_2,0)${,} and negative {elsewhere}.
Since {$L$ is bounded below, all} attractors 
must reside {within this} subspace.
However, {trajectories} cannot {remain confined to it,} 
because arbitrary {points such as} $(x_1,0,x_2,0)$ and 
$(x_1^\prime,0,x_2^\prime,0)$ {cannot be connected through
zero velocity.}
Conceptually, {all points in this} subspace {-- except for} 
three fixed points {--} are linked to {points outside the subspace} via {a} vector field {perpendicular to it.}
{Due to} the monotonically decreasing {nature of the} Lyapunov {function, the flow converges to its minima.}
{Therefore, the only possible attractors are fixed points.}

Next, we {analyze} the stability of the fixed points.
We {consider} the three {typical} fixed points 
\begin{align}
{\mathrm{P}_0:} \, \, & 
x_1=x_2=0, \, v_1=v_2=0, 
\\
{\mathrm{P}_{\pm}:} \, \, & 
x_1=x_2= \pm \sqrt{\kappa / k}, \, v_1=v_2=0.
\end{align}
{To determine} whether {these} are local minima{,}
{we} restrict our analysis to the {two-dimensional} subspace
{because} $L(x_1,v_1,x_2,v_2) \ge L(x_1,0, x_2,0)$.
The first-order partial derivatives {are:}
\begin{eqnarray}
\frac{\partial L}{\partial x_1} & = &  
k x_1^3 - \kappa x_1 - C (x_2 - x_1),\\
\frac{\partial L}{\partial x_2} & = &  
k x_2^3 - \kappa x_2 - C (x_1 - x_2).
\end{eqnarray}
{These} derivatives vanish at {the} fixed points.
{To check} local {minimality, we compute}
the Hessian matrix:
\begin{eqnarray}
    \begin{pmatrix}
     \frac{\partial^2L}{\partial x_1^2}  & \frac{\partial^2L}{\partial x_1\partial x_2}  \\
     \frac{\partial^2L}{\partial x_2\partial x_1} & \frac{\partial^2L}{\partial x_2^2} 
    \end{pmatrix}
    =
    \begin{pmatrix}
     3 k x_1^2 - \kappa + C  & - C  \\
     - C & 3 k x_2^2 - \kappa + C 
    \end{pmatrix}
\nonumber\\
\end{eqnarray}
{Evaluating this at the fixed point $\mathrm{P}_0$, 
we obtain} 
\begin{equation}
 \begin{pmatrix}
     - \kappa + C  & - C  \\
     - C & - \kappa + C 
  \end{pmatrix}
\end{equation}
{with the eigenvalues $-\kappa+2C$ and $-\kappa$.
If $-\kappa+2C < 0$, this matrix} is negative definite, 
{and hence the origin is a local maximum of $L$.
On the other hand, at $\mathrm{P}_\pm$, we obtain}
\begin{equation}
\begin{pmatrix}
     2\kappa + C  & - C  \\
     - C &  2\kappa + C 
 \end{pmatrix}.
\end{equation}
{This matrix is positive definite with eigenvalues $2\kappa$, $2\kappa + 2C$,
making these fixed points global minima of $L$.
In summary, $\mathrm{P}_{\pm}$ are attractors,
and the origin $\mathrm{P}_{0}$ is an unstable fixed point.}

Although {other fixed points may exist}, the global minima are always located at 
{$\mathrm{P}_\pm$, as can be inferred from the inequality:} 
\begin{equation}
 L(x_1,v_1,x_2,v_2) \ge \frac{k}{4}(x_1^4 + x_2^4 ) - \frac{\kappa}{2} (x_1^2+x_2^2),
\end{equation}
{whose global minimum is attaned at $\mathrm{P}_\pm$.} 

\begin{figure}[htbp]
    \includegraphics[width=1.0\linewidth]{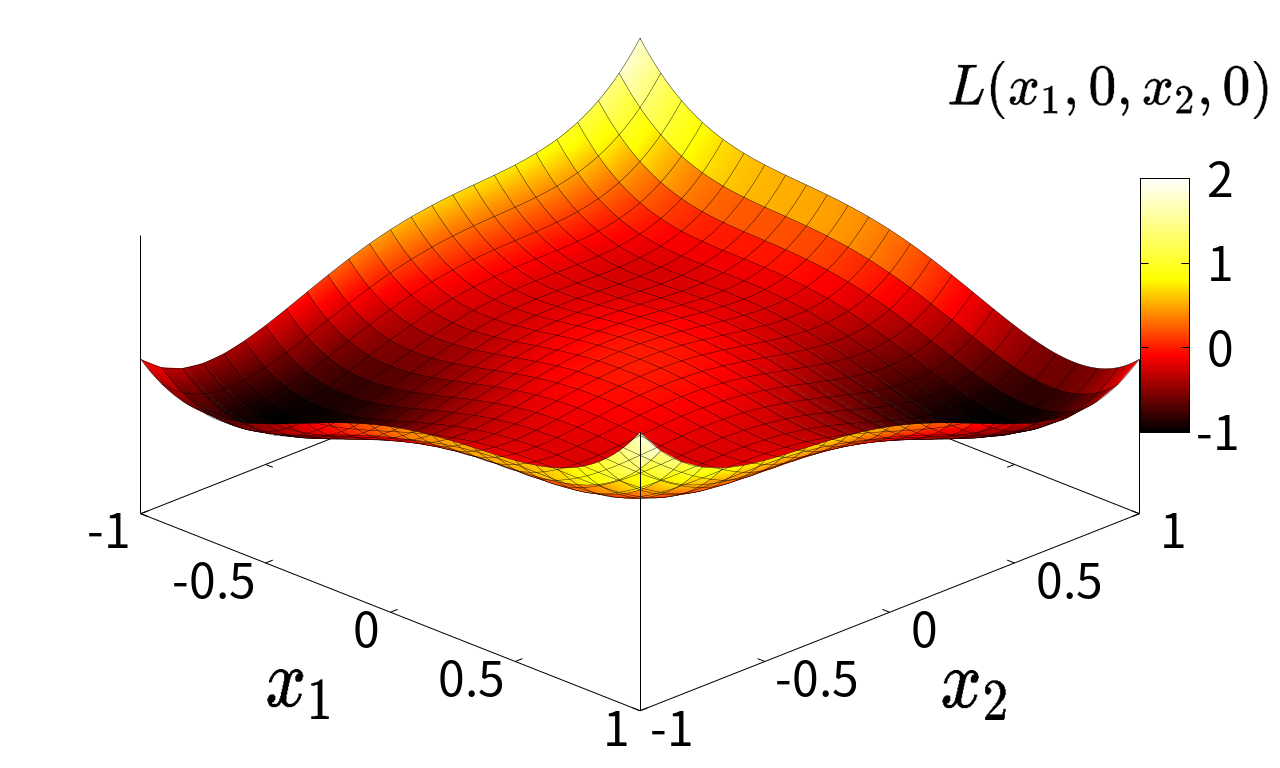}
    \caption{Lyapunov function in {the two-dimensional} subspace.}
    \label{fig:A1}
\end{figure}

\figurename~\ref{fig:A1} shows the {Lyapunov function} $L(x_1,0,x_2,0)$
{for}  $k=8$, $\kappa=4$, {and} $C=1$.
As {expected, the function has} two global minima and one local maximum{,} 
corresponding to the three fixed points.
{This results agrees} with the stability analysis {and supports the existence of
two stable and one unstable fixed points.}
{In addition,} the constant negative first LLE {confirms that no bifurcation occurs
and that only fixed points exist as attractors.}

\renewcommand{\theequation}{B\arabic{equation}}
\setcounter{equation}{0}
\renewcommand{\thefigure}{B\arabic{figure}}
\setcounter{figure}{0}

\section{Reciprocal model of a unit}

\figurename~\ref{fig:B1} provide{s} a conceptual sketch of a 
reciprocal model of a unit, {in contrast to the nonreciprocal case.
This system} corresponds to the differential equations
\begin{widetext}
\begin{equation}
    \dot{x}_{j}=v_{j}, \quad \forall j
\end{equation}
and
\begin{equation}
    \dot{v}_{j}=\left\{\begin{matrix}\begin{array}{l l l}
 - k x_{j}^3 + \kappa x_{j} + C (x_{j-1} - x_{j}) + C (x_{j+1} - x_{j})
    - \gamma v_{j}, & j \neq 1 \text{ and } j \neq n\\ 
  - k x_{j}^3 + \kappa x_{j} + C (x_{n} - x_{j}) + C (x_{j+1} - x_{j})
    - \gamma v_{j}, & j = 1 \\
  - k x_{j}^3 + \kappa x_{j} + C (x_{j-1} - x_{j}) + C (x_{1} - x_{j})
    - \gamma v_{j},& j = n.
\end{array}
\end{matrix}\right.
\end{equation}
\end{widetext}
Although {the unit retains} a ring structure{, it does not exhibit}
nonreciprocal amplification {introduced} by unidirectional couplings.
{As the system exhibits} no amplification{, the dampening} cannot be 
counteracted by the compensation {effects of the ring.
Consequently, the} attractors \/Add{in this system are limited to} fixed points.
{More specifically, 
as demonstrated in Appendix A. 
a Lyapunov function can be constructed for any reciprocal system. 
That analysis shows that trajectories cannot remain confined to 
the subspace where the derivative of the Lyapunov function vanishes.
This guarantees that fixed points are the only admissible attractors in the reciprocal setting.}
\begin{figure}[htbp]
    \includegraphics[width=0.6\linewidth]{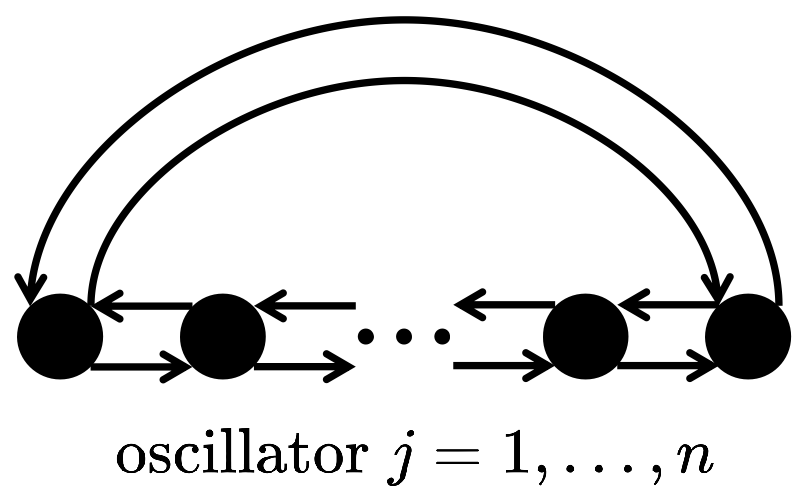}
    \caption{Reciprocal model of a unit. {Solid black circles} 
    represent dissipative double-well Duffing oscillators{, and 
    black arrows indicate} reciprocal couplings $C$.}
    \label{fig:B1}
\end{figure}

\renewcommand{\thefigure}{C\arabic{figure}}
\setcounter{figure}{0}
\section{Criteria for the first LLE}

\begin{figure}[htbp]
 \includegraphics[width=1.0\linewidth]{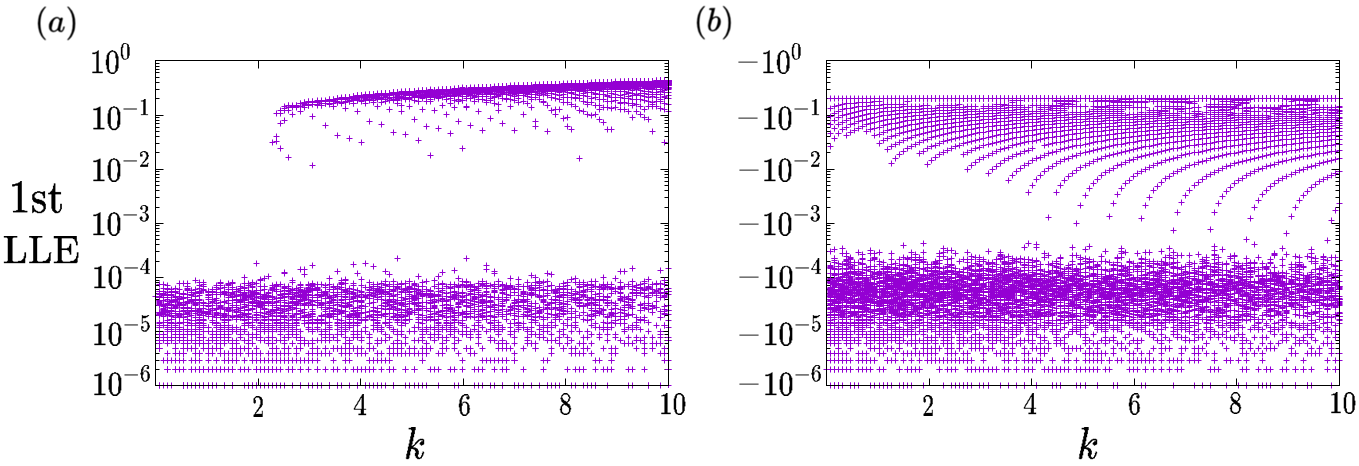}
\caption{{First Lyapunov exponent (LLE) plotted against the} projection of 
{the} $k$-$C$ space onto $k$ axis. 
(a) Positive value{s} of {the} first LLE \textit{vs.} $k$. 
(b) Negative value{s} of {the} first LLE \textit{vs.} $k$. Here, $C^\prime / C=  0.5$. 
}
\label{fig:C1}
\end{figure}

\figurename~\ref{fig:C1}(a) and \figurename~\ref{fig:C1}(b) 
{illustrate the justification for}
the interval in which the values of the first LLE are {considered to be} zero. 
As {shown in both panels,} there are {clear} gaps 
in {the distributions of the first LLE values.}
{We select a threshold} within {this} gap to {define} the bound{s of 
the zero} interval.
The resulting interval is  $(-10^{-3}, 10^{-3})$.
{Any} value larger than $10^{-3}$ ({or} smaller than $-10^{-3}$) {is} regarded 
as {a} positive ({or} negative) first LLE{, respectively}.

\renewcommand{\thefigure}{D\arabic{figure}}
\setcounter{figure}{0}
\section{Robustness of descending pattern}

\begin{figure}[h]
    \includegraphics[width=1.0\linewidth]{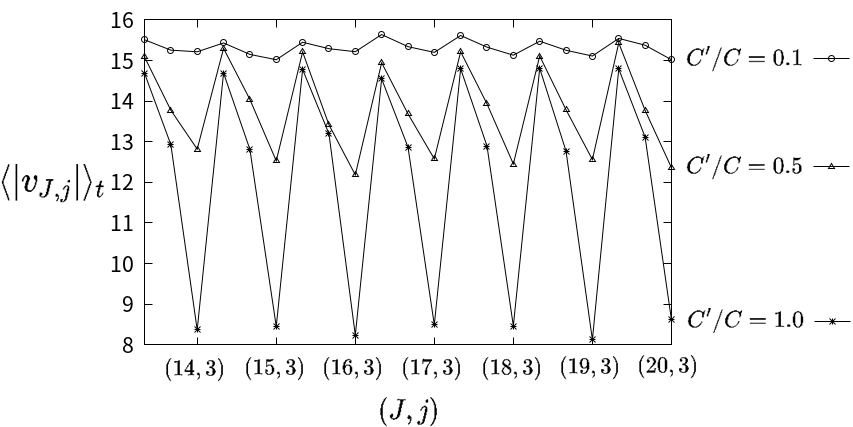}
    \caption{Time-averaged velocity amplitude{s} of oscillators{,} 
    $\langle |v_{J,j}| \rangle_t${, plotted against} oscillator {indices} $(J, j)$.
    The parameters are set {to} $k = 0.3$ {and} $C = 4$.}
    \label{fig:D1}
\end{figure}

\figurename~\ref{fig:D1} presents the descending patterns in the saturation regime for different values of $C^\prime / C$.
Evidently, the slope of the descending pattern within each unit {depends strongly} 
on $C^\prime / C$, {it increases for} larger values of $C^\prime / C$.
In \figurename~\ref{fig:D1}, the oscillator {at position $(J,1)$} receive{s} 
unidirectional {input} from both $(J-1,3)$ and $(J,3)$, 
{resulting in a stronger amplification}  than {that of the other oscillators}.
{Consequently, the} descending pattern can be {interpreted} as an {emergent} adaptation of the system to {velocity amplitude saturation.}
When {$C^\prime / C \to 0$,  the coupling} from $(J,3)$ to $(J,1)$ becomes negligible, {effectively} restoring 
translational symmetry among {the} oscillators.
{This leads}  to nearly uniform saturated amplitudes {across each unit}.
Conversely, {for} large values of $C^\prime / C${, this} translational symmetry 
{is broken} at $(J,1)$, enforcing a descending distribution of saturated velocity amplitudes 
{within} each unit.
This descending pattern {-- shaped} by the interplay between coupling structure and 
saturation behavior {-- remains} robust over a broad range of {parameter values}.

\renewcommand{\thefigure}{E\arabic{figure}}
\setcounter{figure}{0}
\section{Coexistence of attractors}

{When multiple attractors exist in phase space, the} choice of initial conditions can 
{significantly affect the long-term dynamics.}
{\figurename~\ref{fig:E1},
based on simulations from} 1,500 different initial conditions, 
reveals four distinct values of the first LLE, each corresponding to a different attractor.

\begin{figure}[htbp]
    \includegraphics[width=1.0\linewidth]{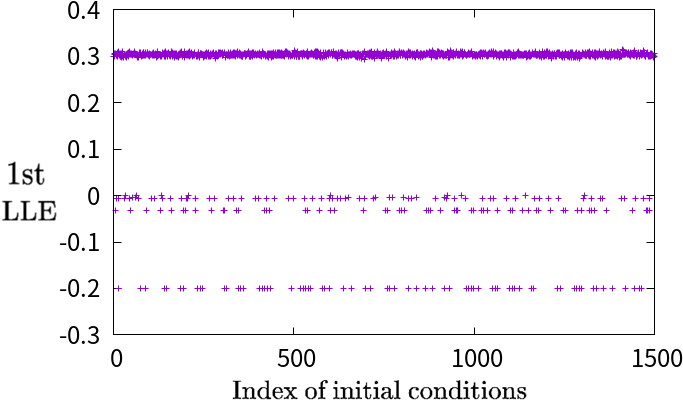}
    \caption{Dependence of first {Lyapunov exponent (LLE)} on different initial conditions. 
    The horizontal axis indicates the index of the initial conditions, {and} the vertical axis {shows} the corresponding values of the first LLE. 
    The parameters are set {to} $k = 6$, $C = 3$, and $C^\prime/C = 0.5$.}
    \label{fig:E1}
\end{figure}


\input{zhao2025nonreciprocal_rev2.bbl}

\end{document}

%% file: zhao2025nonreciprocal_rev2.bbl
%